\documentclass[twocolumn]{article}
\usepackage[utf8]{inputenc}
\usepackage[big,online]{dgruyter}   %values: small,big | online,print,work
\usepackage{lmodern}
\usepackage{mathbbol}
\usepackage{amsmath,amssymb,bm}
\usepackage{graphicx}
\usepackage{microtype}
\usepackage[numbers,square,sort&compress]{natbib}
\usepackage[usenames,dvipsnames]{xcolor}
\hypersetup{colorlinks,allcolors=blue}

\def\equationautorefname~#1\null{Eq.~(#1)\null}

\newcommand{\Int}{\int\limits}
\newcommand{\Sum}{\sum\limits}

\newcommand{\dd}{\mathrm{d}}
\renewcommand{\Im}{\operatorname{Im}}

\newcommand{\ff}{\hat{\mathbf{f}}_{\lambda}(\mathbf{r},\omega)}
\newcommand{\ffdag}{\hat{\mathbf{f}}_{\lambda}^{\dagger}(\mathbf{r},\omega)}
\newcommand{\ffp}{\hat{\mathbf{f}}_{\lambda'}(\mathbf{r'},\omega')}
\newcommand{\ffpdag}{\hat{\mathbf{f}}_{\lambda'}^{\dagger}(\mathbf{r'},\omega')}

\begin{document}

%%%--------------------------------------------%%%
%   \articletype{Research Article}
%   \received{Month   DD, YYYY}
%   \revised{Month    DD, YYYY}
%   \accepted{Month   DD, YYYY}
%   \journalname{Nanophotonics}
%   \journalyear{YYYY}
%   \journalvolume{XX}
%   \journalissue{X}
%   \startpage{1}
%   \aop
%   \DOI{10.1515/sample-YYYY-XXXX}
%%%--------------------------------------------%%%

\title{Macroscopic QED for quantum nanophotonics: Emitter-centered modes as a minimal basis for multi-emitter problems}
\runningtitle{Macroscopic QED for quantum nanophotonics}

%\ use * to mark the author as the corresponding author
\author*[1]{Johannes Feist}
\author[2]{Antonio I. Fernández-Domínguez}
\author*[3]{Francisco J. García-Vidal}
\runningauthor{J.~Feist et al.}
\affil[1]{\protect\raggedright
Departamento de Física Teórica de la Materia Condensada and Condensed Matter
Physics Center (IFIMAC), Universidad Autónoma de Madrid, E-28049 Madrid, Spain,
e-mail: johannes.feist@uam.es}
\affil[2]{\protect\raggedright
Departamento de Física Teórica de la Materia Condensada and Condensed Matter
Physics Center (IFIMAC), Universidad Autónoma de Madrid, E-28049 Madrid, Spain,
e-mail: a.fernandez-dominguez@uam.es}
\affil[3]{\protect\raggedright
Departamento de Física Teórica de la Materia Condensada and Condensed Matter
Physics Center (IFIMAC), Universidad Autónoma de Madrid, E-28049 Madrid, Spain,
and
Donostia International Physics Center (DIPC), E-20018 Donostia/San Sebastian, Spain,
email: fj.garcia@uam.es}

\abstract{We present an overview of the framework of macroscopic
quantum electrodynamics from a quantum nanophotonics perspective.
Particularly, we focus our attention on three aspects of the
theory which are crucial for the description of quantum optical
phenomena in nanophotonic structures. First, we review the
light-matter interaction Hamiltonian itself, with special emphasis
on its gauge independence and the minimal and multipolar coupling
schemes. Second, we discuss the treatment of the external pumping
of quantum-optical systems by classical electromagnetic fields. Third,
we introduce an exact, complete and minimal basis for the field
quantization in multi-emitter configurations, which is based on
the so-called emitter-centered modes. Finally, we illustrate this
quantization approach in a particular hybrid metallodielectric
geometry: two quantum emitters placed in the vicinity of a dimer
of Ag nanospheres embedded in a SiN microdisk.}


\maketitle

\section{Introduction}
In principle, quantum electrodynamics (QED) provides an ``exact'' approach for
treating electromagnetic (EM) fields, charged particles, and their interactions,
within a full quantum field theory where both matter and light are second
quantized (i.e., both photons and matter particles can be created and
annihilated). However, this approach is not very useful for the treatment of many
effects of interest in fields such as (nano)photonics and quantum optics, which
take place at ``low'' energies (essentially, below the rest mass energy of
electrons), where matter constituents are stable and neither created nor
destroyed, and additionally, there are often ``macroscopic'' structures such as
mirrors, photonic crystals, metallic nanoparticles etc.\ involved. Due to the
large number of material particles (on the order of the Avogadro
constant, $\approx 6\cdot 10^{23}$), it then becomes unthinkable to treat the
electrons and nuclei in these structures individually. At the same time, a
sufficiently accurate description of these structures is usually given by the
macroscopic Maxwell equations, in which the material response is described by
the constitutive relations of macroscopic electromagnetism. In many situations,
it is then desired to describe the interactions between light and matter in a
setup where there are one or a few microscopic ``quantum emitters'' (such as
atoms, molecules, quantum dots, etc.), and additionally a ``macroscopic''
material structure whose linear response determines the local modes of the EM
field interacting with the quantum emitter(s).

The quantization of the EM field in such arbitrary material environments, i.e.,
the construction of a second quantized basis for the medium-assisted EM field
that takes into account the presence of the ``macroscopic'' material structure,
is a longstanding problem in quantum electrodynamics. The most immediate
strategy is to calculate the (classical) EM modes of a structure and to quantize
them by normalizing their stored energy to that of a single photon at the mode
frequency, $\hbar\omega$~\cite{Loudon2000}. However, an important point to
remember here is that, even for lossless materials, electromagnetic modes always
form a continuum in frequency, i.e., there exist modes at any positive frequency
$\omega$. Consequently, there are in general no truly bound EM modes, and what
is normally thought of as an isolated ``cavity mode'' is more correctly
described as a resonance embedded in the continuum, i.e., a quasi-bound state
that decays over time through emission of radiation. An interesting exception
here are guided modes in systems with translational invariance (i.e., where
momentum in one or more dimensions is conserved), as modes lying outside the
light cone $\omega = c k$ then do not couple to free-space
radiation~\cite{Abujetas2019}. An additional exception are ``bound states in the
continuum''~\cite{Hsu2016}, which arise due to destructive interference between
different resonances.

As a further obstacle to a straightforward quantization strategy as described
above, the response functions describing material structures are necessarily
dissipative due to causality (as encoded in, e.g., the Kramers-Kronig
relations). When these losses cannot be neglected, quantization is complicated
even further by the difficulty to define the energy density of the EM field
inside the lossy material~\cite{Loudon1970, Maier2007, Vazquez-Lozano2018}.

Given all the points above, it is not surprising that there are many different
approaches to quantizing EM modes in lossy material systems~\cite{Imamoglu1994,
Garraway1997Decay, Garraway1997Nonperturbative, Dalton2001, Waks2010,
Gonzalez-Tudela2013, Alpeggiani2014, Abujetas2019, Franke2019, Lentrodt2020}. In
the following, we give a concise overview of a particularly powerful formal
approach that resolves these problems, called macroscopic quantum
electrodynamics (macroscopic QED)~\cite{Fano1956, Hopfield1958, Huttner1992,
Dung1998, Scheel1998, Buhmann2007Dispersion, Scheel2008, Buhmann2012I,
Buhmann2012II}. While there are excellent reviews of this general framework
available (e.g.,~\cite{Scheel2008, Buhmann2012I}), we focus on its application
in the context of quantum nanophotonics and strong light-matter coupling. In
particular, we discuss the implications and lessons that can be taken from this
approach on gauge independence and, in particular, the role of the so-called
dipole self-energy term in the light-matter interaction in the
Power-Zienau-Woolley gauge, which has been the subject of some recent
controversy~\cite{Vukics2014, DeBernardis2018Breakdown, DeBernardis2018Cavity,
Rokaj2018, Andrews2018, Vukics2018, Rousseau2018, SanchezMunoz2018Resolution,
Schafer2018, Stokes2019, Galego2019, Taylor2020, DiStefano2019}. We then review
in detail a somewhat non-standard formulation of macroscopic QED that allows one
to construct a minimal quantized basis for the EM field interacting with a
collection of multiple quantum emitters. This approach was first introduced
in~\cite{Buhmann2008Casimir}, and subsequently rediscovered independently by
several other groups~\cite{Hummer2013, Rousseaux2016, Dzsotjan2016,
Castellini2018, Varguet2019Non-hermitian}. The fact that this very useful
approach has been reinvented by different researchers over the past decade or so
partially motivates the current article, which intends to give a concise and
accessible overview, and presents some explicit relations that have (to our
knowledge) not been published before. We also note that with ``minimal basis''
we are here referring to a minimal \emph{complete} basis for the medium-assisted
EM field, i.e., this basis contains all the information about the material
structure playing the role of the cavity or antenna, and no approximations are
made in obtaining it. This then makes it appropriate to serve as a starting
point for either numerical treatments~\cite{Sanchez-Barquilla2020, Zhao2020}, or
for deriving simpler models where, e.g., the full EM spectrum is described by a
few lossy modes~\cite{Medina2020}.

In the final part of the article, we then present an application of the
formalism to a specific problem, a hybrid dielectric-plasmonic
structure ~\cite{Peng2017,Gurlek2018,Franke2019}.
Particularly, we consider a dimer of metallic nanospheres placed within a dielectric microdisk,
a geometry that is similar to that considered in~\cite{Doeleman2016}.

\section{Theory}\label{sec:theory}

Macroscopic QED provides a recipe for quantizing the EM field in any geometry,
including with lossy materials. One particularly appealing aspect is that the
full information about the quantized EM field is finally encoded in the
(classical) electromagnetic dyadic Green's function
$\mathbf{G}(\mathbf{r},\mathbf{r}',\omega)$. While there are several ways to
derive the general formulation (see, e.g.,~\cite{Scheel2008} for a discussion of
various approaches), a conceptually simple way to understand the framework is to
represent the material structures through a collection of fictitious harmonic
oscillators coupled to the free-space EM field (which itself corresponds to a
collection of harmonic oscillators~\cite{Grynberg2010}). Formally diagonalizing
this system of coupled harmonic oscillators leads to a form where the linear
response of the medium can be expressed through the coupling between the
material oscillators and the EM field. The end result is that the fully
quantized medium-supported EM field is represented by an infinite set of bosonic
modes (labelled with index $\lambda$, see below), defined at each point in space and each frequency, $\ff$, which act as
sources for the EM field through the classical Green's function. These modes are
called ``polaritonic'' as they represent mixed light-matter
excitations~\cite{Hopfield1958}. While this is a completely general approach for
quantizing the EM field in arbitrary structures, it cannot be used ``directly''
in practice due to the extremely large number of modes that describe the EM
field (a vectorial-valued 4-dimensional continuum). Most uses of macroscopic QED
thus apply this formalism to derive expressions where the explicit operators
$\ff$ have been eliminated, e.g., through adiabatic elimination, perturbation
theory, or the use of Laplace transform techniques~\cite{Wubs2004, Scheel2008,
Yao2009, Buhmann2012I, Buhmann2012II, Delga2014, Asenjo-Garcia2017}. In particular,
macroscopic QED has been widely used in the context of dispersion forces, such
as described by Casimir and Casimir-Polder
potentials~\cite{Buhmann2012I,Buhmann2012II}.

% , i.e., when the imaginary part
% of the refractive index $n(\mathbf{r},\omega) =
% \sqrt{\varepsilon(\mathbf{r},\omega) \mu(\mathbf{r},\omega)}$ cannot be
% neglected, there are no bound modes even in a (fictitious) box

% As discussed above, we will now derive a somewhat non-standard formulation of
% macroscopic QED, obtained through a linear transformation of the fundamental
% basis $\ff$. In this formulation, only $N$ ``bright'' or ``emitter-centered''
% modes $\hat{C}_i(\omega)$ are required at each frequency to fully represent the
% interaction between the EM field and $N$ (dipolar) emitters.

\subsection{Minimal coupling}
In the following, we represent a short overview of the general theory of
light-matter interactions in the framework of macroscopic QED. Since full
details can be found in the literature~\cite{Scheel2008,Buhmann2012I}, we do not
attempt to make this a fully self-contained overview, but rather highlight and
discuss some aspects that are not within the traditional focus of the theory, in
particular in the context of quantum nanophotonics.

The first step in the application of macroscopic QED is the separation of all
matter present in the system to be treated into two distinct groups: one is the
macroscopic structure (e.g., a cavity, plasmonic nanoantenna, photonic crystal,
\ldots) that will be described through the constitutive relations of
electromagnetism, while the other are the microscopic objects (atoms, molecules,
quantum dots, \ldots) that are described as a collection of charged particles
(which can then be approximated at various levels, e.g., as two-level systems).
This separation constitutes the basic approximation inherent in the approach,
and relies on the assumptions that macroscopic EM is valid for the material
structure (the medium) and its interaction with the charged particles. While
this is often an excellent approximation, some care has to be taken for
separations in the sub-nanometer range, where the atomic structure of the
material can have a significant influence~\cite{Savage2012, Zhang2014,
Aguirregabiria2018, Sinha-Roy2017, Zhang2017, Chen2018}

For simplicity, we assume that the medium response is local and isotropic in
space, such that it can be encoded in the position- and frequency-dependent
scalar relative permittivity $\epsilon(\mathbf{r},\omega)$ and relative
permeability $\mu(\mathbf{r},\omega)$ that describe the local matter
polarization and magnetization induced by external EM fields\footnote{Note that
since the response functions are considered time-independent, effects due to the
motion of the structure, such as in cavity optomechanics~\cite{Aspelmeyer2014},
cannot be treated without further modifications.}. The extension of the
quantization scheme to nonlocal response functions can be found
in~\cite{Scheel2008}. We directly give the Hamiltonian $\mathcal{H} =
\mathcal{H}_A + \mathcal{H}_F + \mathcal{H}_{AF}$ within the minimal coupling
scheme~\cite{Scheel2008,Buhmann2012I}
\begin{subequations}\label{eq:H_MQED_mincoup_full}
\begin{align}
    \mathcal{H}_A &= \sum_\alpha \frac{\hat{\mathbf{p}}_\alpha^2}{2m_\alpha} + \sum_\alpha \sum_{\beta>\alpha} \frac{q_\alpha q_\beta}{4\pi\varepsilon_0 |\mathbf{r}_\alpha - \mathbf{r}_\beta|}, \label{eq:H_A_mincoup}\\
    \mathcal{H}_F &= \sum_\lambda \Int_0^\infty \dd\omega \Int \dd^3\mathbf{r}\, \hbar\omega\,\ffdag \ff, \label{eq:H_F_mincoup}\\
    \mathcal{H}_{AF} &= \sum_\alpha \left[ q_\alpha \hat{\phi}(\hat{\mathbf{r}}_\alpha) - \frac{q_\alpha}{m_\alpha} \hat{\mathbf{p}}_\alpha \cdot\hat{\mathbf{A}}(\hat{\mathbf{r}}_\alpha) + \frac{q_\alpha^2}{2m_\alpha} \hat{\mathbf{A}}^2(\hat{\mathbf{r}}_\alpha) \right]. \label{eq:H_AF_mincoup}
\end{align}
\end{subequations}
Here, the ``atomic'' Hamiltonian $\mathcal{H}_A$ describes a (nonrelativistic)
collection of point particles with position and momentum operators
$\hat{\mathbf{r}}_\alpha$ and $\hat{\mathbf{p}}_\alpha$, and charges and masses
$q_\alpha$ and $m_\alpha$. The field Hamiltonian $\mathcal{H}_F$ is expressed in
terms of the bosonic operators $\ff$ discussed above, which obey the commutation
relations
\begin{subequations}\label{eq:f_comm}
\begin{align}
    \left[\ff,\, \ffpdag\right] &= \delta_{\lambda\lambda'} \boldsymbol{\delta}(\mathbf{r}-\mathbf{r}') \delta(\omega-\omega'),\\
    \left[\ff,\, \ffp\right] &= \left[\ffdag,\, \ffpdag\right] = \boldsymbol{0},
\end{align}
\end{subequations}
where $\boldsymbol{\delta}(\mathbf{r}-\mathbf{r}') =
\delta(\mathbf{r}-\mathbf{r}') \boldsymbol{1}$, and $\boldsymbol{1}$ and
$\boldsymbol{0}$ are the Cartesian ($3\times3$) identity and zero tensors,
respectively. The index $\lambda \in \{e,m\}$ labels the electric and magnetic
contributions, with the magnetic contribution disappearing if
$\mu(\mathbf{r},\omega) = 1$ everywhere in space. The particle-field interaction
Hamiltonian $\mathcal{H}_{AF}$ contains the interaction of the charges both with
the electrostatic potential $\hat{\phi}(\mathbf{r})$ and the vector potential
$\hat{\mathbf{A}}(\mathbf{r})$, both of which can be expressed through the
fundamental operators $\ff$~\cite{Scheel2008,Buhmann2012I}. Note that in the
above, we have neglected magnetic interactions due to particle spin. We
explicitly point out that although we work in Coulomb gauge, $\nabla \cdot
\hat{\mathbf{A}}(\mathbf{r}) = 0$, the electrostatic potential
$\hat{\phi}(\mathbf{r})$ is in general nonzero due to the presence of the
macroscopic material structure, which also implies that the name
``$\mathbf{p}\cdot \mathbf{A}$-gauge'', which is sometimes employed for the
minimal coupling scheme, is misleading in the presence of material bodies.

The light-matter interaction Hamiltonian can be simplified in the long-wavelength or
dipole approximation, i.e., if we assume that the charged particles are
sufficiently close to each other compared to the spatial scale of local field
variations that a lowest-order approximation of the positions of the charges
relative to their center of mass position $\hat{\mathbf{r}}_i$ is valid. For an
overall neutral collection of charges, this leads to
\begin{equation}\label{eq:H_AF_mincoup_dipole}
    \mathcal{H}_{AF} \approx -\hat{\mathbf{d}}\cdot \hat{\mathbf{E}}^\parallel(\hat{\mathbf{r}}_i)
    - \sum_\alpha \frac{q_\alpha}{m_\alpha} \hat{\bar{\mathbf{p}}}_\alpha \cdot\hat{\mathbf{A}}(\hat{\mathbf{r}}_i)
    + \sum_\alpha \frac{q_\alpha^2}{2m_\alpha} \hat{\mathbf{A}}^2(\hat{\mathbf{r}}_i),
\end{equation}
where $\hat{\mathbf{d}} = \sum_\alpha q_\alpha \hat{\bar{\mathbf{r}}}_\alpha$ is
the electric dipole operator of the collection of charges, while
$\hat{\bar{\mathbf{r}}}_\alpha$ and $\hat{\bar{\mathbf{p}}}_\alpha$ are the
position and momentum operators in the center-of-mass frame of the charge
collection\footnote{While we here assumed a single emitter (i.e., a collection
of close-by charged particles), the extension of the above formula to multiple
emitters is trivial. One important aspect to note is that the electrostatic
Coulomb interaction between charges (second term in \autoref{eq:H_A_mincoup}) is
still present, i.e., there are direct instantaneous Coulomb interactions between
the charges in different emitters.}. \autoref{eq:H_AF_mincoup_dipole} explicitly
shows that in the presence of material bodies, the emitter-field interaction has
two contributions, one from longitudinal (electrostatic) fields due to the
Coulomb interaction with charges in the macroscopic body, and one from
transverse fields (described by the vector potential). The relevant fields are
given by~\cite{Buhmann2012I}
\begin{subequations}\label{eq:E_long_A_MQED}
\begin{align}
    \hat{\mathbf{E}}^\parallel(\hat{\mathbf{r}}) &= \Sum_\lambda \Int_0^\infty \dd\omega \Int\dd^3\mathbf{r}'\, {}^\parallel\mathbf{G}_\lambda(\mathbf{r},\mathbf{r}',\omega) \cdot \hat{\mathbf{f}}_\lambda(\mathbf{r}',\omega) + \mathrm{H.c.} \label{eq:E_long_MQED},\\
    \hat{\mathbf{A}}(\hat{\mathbf{r}}) &= \Sum_\lambda \Int_0^\infty \frac{\dd\omega}{i\omega} \Int\dd^3\mathbf{r}'\, {}^\perp\mathbf{G}_\lambda(\mathbf{r},\mathbf{r}',\omega) \cdot \hat{\mathbf{f}}_\lambda(\mathbf{r}',\omega) + \mathrm{H.c.} \label{eq:A_MQED},
\end{align}
\end{subequations}
where the longitudinal and transverse components of a tensor
$\mathbf{T}(\mathbf{r},\mathbf{r}')$ are given by
\begin{equation}
    {}^{\parallel/\perp}\mathbf{T}(\mathbf{r},\mathbf{r}') = \Int\dd^3\mathbf{s}\,\boldsymbol{\delta}^{\parallel/\perp}(\mathbf{r}-\mathbf{s}) \mathbf{T}(\mathbf{s},\mathbf{r}'),
\end{equation}
with $\boldsymbol{\delta}^{\parallel/\perp}(\mathbf{r}-\mathbf{s})$ the standard
longitudinal or transverse delta function in 3D space. The functions
$\mathbf{G}_\lambda(\mathbf{r},\mathbf{r}',\omega)$ are related to the dyadic
Green's function $\mathbf{G}(\mathbf{r},\mathbf{r'},\omega)$ through
\begin{subequations}
\begin{gather}
\mathbf{G}_e(\mathbf{r},\mathbf{r}',\omega) = i \frac{\omega^2}{c^2} \sqrt{\frac{\hbar}{\pi\epsilon_0} \Im\epsilon(\mathbf{r'},\omega)} \mathbf{G}(\mathbf{r},\mathbf{r'},\omega),\\
\mathbf{G}_m(\mathbf{r},\mathbf{r}',\omega) = i \frac{\omega}{c} \sqrt{\frac{-\hbar}{\pi\epsilon_0} \Im\mu^{-1}(\mathbf{r'},\omega)} [\mathbf{\nabla}' \times \mathbf{G}(\mathbf{r'},\mathbf{r},\omega)]^T.
\end{gather}
\end{subequations}
We note that in the derivation leading to the above expressions, it is assumed
that $\Im\epsilon(\mathbf{r},\omega) > 0$ and $\Im\mu(\mathbf{r},\omega)>0$
for all $\mathbf{r}$, i.e., that the materials are lossy everywhere in space.
The limiting case of zero losses in some regions of space (e.g., in free space)
is only taken at the very end of the calculation. We will see that in the
reformulation in terms of emitter-centered modes,
\autoref{sec:emitter_centered_modes},
$\mathbf{G}_\lambda(\mathbf{r},\mathbf{r}',\omega)$ disappears from the
formalism relatively early, and only the ``normal'' Green's function
$\mathbf{G}(\mathbf{r},\mathbf{r'},\omega)$ is needed (for which the limit is
straightforward).

As mentioned above, the electrostatic contribution is not present in free space,
and in the literature it is often assumed that any abstract ``cavity mode''
corresponds to a purely transverse EM field. This is a good approximation for
emitters that are far enough away from the material, e.g., in ``large''
(typically dielectric) structures such as Fabry-Perot planar microcavities,
photonic crystals, micropillar resonators, etc.~\cite{Sanvitto2016}, but can
break down otherwise. In general, this happens for coupling to evanescent
fields~\cite{Petersen2014}, and in particular when sub-wavelength confinement is
used to generate extremely small effective mode volumes, such as in
plasmonic~\cite{Fernandez-Dominguez2017,Baumberg2019} or phonon-polaritonic
systems~\cite{Foteinopoulou2019,Gubbin2020}. This observation is particularly
relevant as such sub-wavelength confinement is the only possible strategy for
obtaining large enough light-matter coupling strengths to approach the
single-emitter strong-coupling regime at room temperature~\cite{Santhosh2016,
Chikkaraddy2016, Gross2018, Ojambati2019}. For sub-wavelength separations, it is
well-known that the Green's function is dominated by longitudinal components,
while transverse components can be neglected~\cite{Buhmann2007Thesis}. In this
\emph{quasistatic} approximation, we thus have $\mathbf{\hat{A}} \approx 0$,
cf.~\autoref{eq:E_long_A_MQED}, and the light-matter interactions are all due to
electrostatic (or Coulomb) interactions, even within the minimal coupling
scheme~\cite{Galego2019}\footnote{It could then be discussed what the field
modes should be called in the limit when they contain negligible contributions
from propagating photon modes. However, since \emph{all} EM modes that are not
just freely propagating photons will always have a somewhat mixed light-matter
character, and since these modes always solve the macroscopic Maxwell equations,
they are conventionally referred to as ``light'', ``EM'', or ``photon'' modes.
It is thus important to remember that this does not imply that they are simply
modes of the transverse EM field.}. Conversely, due to the strongly
sub-wavelength field confinement, the long-wavelength or dipole approximation is
not necessarily appropriate and an accurate description requires either the
direct use of the expression in terms of the electrostatic
potential~\cite{Neuman2018Coupling}, or the inclusion of higher-order terms in
the interaction~\cite{Cuartero-Gonzalez2018}. As we have previously pointed
out~\cite{Galego2019}, doing so also resolves the formal lack of a ground state
when the computational box is made too large and the dipole approximation is
used~\cite{Rokaj2018,Schafer2018}.

\subsection{Multipolar coupling}
We now discuss the Power-Zienau-Woolley (PZW) gauge
transformation~\cite{Power1959, Woolley1971,Woolley2020}, which is used to
switch from the minimal coupling scheme discussed up to now to the so-called
multipolar coupling scheme, which will then in turn form the basis for the
emitter-centered modes we discuss later. This scheme has several advantageous
properties: it expresses all light-matter interactions through the fields
$\mathbf{E}$ and $\mathbf{B}$ directly, without needing to distinguish between
longitudinal and transverse fields, and allows a systematic expansion of the
field-emitter interactions in terms of multipole moments. Additionally, in the
multi-emitter case, it also removes direct Coulomb interactions between charges
in different emitters, which instead become mediated through the EM fields. This
property makes it easier to explicitly verify and guarantee that causality is
not violated through faster-than-light interactions. We only point out and
discuss some specific relevant results here, and again refer the reader to the
literature for full details~\cite{Scheel2008,Buhmann2012I}. The PZW
transformation is carried out by the unitary transformation operator
\begin{equation}\label{eq:U_PZW}
    \hat{U} = \exp\left[ \frac{i}{\hbar} \int\dd^3\mathbf{r}\, \sum_i \hat{\mathbf{P}}_i \cdot \hat{\mathbf{A}} \right]
\end{equation}
where $\hat{\mathbf{P}}_i = \sum_{\alpha\in i} q_\alpha \hat{\bar{\mathbf{r}}}
\int_0^1 \dd\sigma \delta(\mathbf{r} - \hat{\mathbf{r}}_i - \sigma
\hat{\bar{\mathbf{r}}})$ is the polarization operator of emitter $i$, and we
have here explicitly grouped the charges into several emitters, i.e., distinct
(non-overlapping) collections of charges. Since this is a unitary
transformation, physical results are unaffected in principle, although the
convergence behavior with respect to different approximations can be quite
different~\cite{DeBernardis2018Breakdown,DiStefano2019}.

Applying the transformation above gives the new operators $\hat{O}' = \hat{U} \hat{O}
\hat{U}^\dagger$. Expressing the Hamiltonian \autoref{eq:H_MQED_mincoup_full} in
terms of these new operators then gives the multipolar coupling form. The effect
can be summarized by noting that the operators $\hat{\mathbf{A}}$ and
$\hat{\mathbf{r}}_\alpha$ are unchanged, while their canonically conjugate
momenta $\hat{\mathbf{\Pi}}$ and $\hat{\mathbf{p}}_\alpha$ are not. In light of
the discussion of longitudinal (electrostatic) versus transverse interactions,
it is interesting to point out that the electrostatic potential
$\hat{\phi}(\mathbf{r})$ is also unaffected, i.e., in the quasistatic limit, the
PZW transformation has no effect on the emitter-field interaction and
discussions of gauge dependence for this specific case become somewhat
irrelevant. However, both the bare-emitter and the bare-field Hamiltonian are
changed, as $\ff' \neq \ff$, i.e., the separation into emitter and field
variables is different than in the minimal coupling scheme. We here directly
show the multipolar coupling Hamiltonian after additionally neglecting
interactions containing the magnetic field. This leads to
\begin{subequations}\label{eq:H_MQED_multipolar_full}
\begin{align}
    \mathcal{H}_A &= \sum_i \left[\sum_{\alpha\in i} \frac{\hat{\mathbf{p}}_\alpha^2}{2m_\alpha} + \frac{1}{2\varepsilon_0} \int\dd^3\mathbf{r} \, \hat{\mathbf{P}}_i^2(\mathbf{r}) \right], \label{eq:H_A_multipolar}\\
    \mathcal{H}_F &= \sum_\lambda \Int_0^\infty \dd\omega \Int \dd^3\mathbf{r}\, \hbar\omega\,\ffdag \ff, \label{eq:H_F_multipolar}\\
    \mathcal{H}_{AF} &= -\sum_i \int\dd^3\mathbf{r} \, \hat{\mathbf{P}}_i(\mathbf{r}) \cdot \hat{\mathbf{E}}(\mathbf{r}), \label{eq:H_AF_multipolar}
\end{align}
\end{subequations}
where all operators are their PZW-transformed (primed) versions, but we have not
included explicit primes for simplicity. In the long-wavelength limit,
\autoref{eq:H_AF_multipolar} becomes simply
\begin{equation}
\mathcal{H}_{AF} = -\sum_i \hat{\mathbf{d}}_i \cdot \hat{\mathbf{E}}(\hat{\mathbf{r}}_i),\label{eq:H_AF_multipolar_dipole}
\end{equation}
i.e., all field-emitter interactions are expressed through the dipolar coupling
term, with the electric field operator given explicitly by
\begin{equation}
    \hat{\mathbf{E}}(\mathbf{r}) = \Sum_\lambda \Int_0^\infty \dd\omega \Int\dd^3\mathbf{r}'\, \mathbf{G}_\lambda(\mathbf{r},\mathbf{r}',\omega) \cdot \hat{\mathbf{f}}_\lambda(\mathbf{r}',\omega) + \mathrm{H.c.} \label{eq:E_MQED}.
\end{equation}

We note that the form of the bare-emitter Hamiltonian $\mathcal{H}_A = \sum_i
H_i$ [\autoref{eq:H_A_multipolar}] under multipolar coupling is changed compared to
the minimal coupling picture, and in particular can be rewritten as
\begin{multline}\label{eq:H_A_multipolar_coulomb}
    H_i = \sum_{\alpha\in i} \frac{\hat{\mathbf{p}}_\alpha^2}{2m_\alpha} +
    \sum_{\alpha,\beta \in i} \frac{q_\alpha q_\beta}{8\pi\varepsilon_0 |\mathbf{r}_\alpha - \mathbf{r}_\beta|} \\
    + \frac{1}{2\varepsilon_0} \int\dd^3\mathbf{r} \, \left[\hat{\mathbf{P}}_i^\perp(\mathbf{r})\right]^2,
\end{multline}
which makes explicit the fact that the emitter Hamiltonian in the multipolar
gauge is equivalent to the emitter Hamiltonian in minimal coupling plus a term
containing the transverse polarization only. In order to arrive at this form, we
have used that the Coulomb interaction can be rewritten as an integral over the
longitudinal polarization. The transverse part of the polarization in
\autoref{eq:H_A_multipolar_coulomb} corresponds to the so-called dipole
self-energy term~\cite{Rokaj2018}. When a single- or few-mode approximation of
the EM field is performed \emph{before} doing the PZW transformation, this term
depends on the square of the mode-emitter coupling strength, but not on any
photonic operator. However, when all modes of the EM field are included, as
implicitly done here and as motivated by the fact that the term is not
mode-selective (it does not depend on any EM field operator), it is seen
directly that this term becomes completely independent of any characteristics of
the surrounding material structure, i.e., it cannot be modified by changing the
environment that the emitter is located in. Instead, the bare-emitter
Hamiltonian in the multipolar approach is simply slightly different than under
minimal coupling. This calls into question claims that this term should be
included in simulations of strongly coupled light-matter
systems~\cite{Rokaj2018,Schafer2018}. However, it should be mentioned that a
similar term can arise as if the environment-mediated electrostatic interactions
are taken into account explicitly instead of through the quantized modes, and
one (or some) of the EM modes are additionally treated explicitly. The action of
these modes on the emitters then has to be subtracted from the electrostatic
interaction to avoid double-counting them~\cite{DeBernardis2018Cavity}.

\subsection{External (classical) fields}
Adapting an argument from Ref.~\cite{Sanchez-Barquilla2020}, here we show that
macroscopic QED also enables a straightforward treatment of external incoming
electromagnetic fields, in particular for the experimentally most relevant case
of a classical laser pulse. Assuming that the incoming laser field at the
initial time $t=0$ has not yet interacted with the emitters (i.e., it describes
a pulse localized in space in a region far away from the emitters), it can
simply be described by a product of coherent states of the EM modes for the
initial wave function, $|\psi(0)\rangle = \prod_n|\alpha_n(0)\rangle = \prod_n
e^{\alpha_n(0) a_n^\dagger - \alpha_n^*(0) a_n}|0\rangle$, where the index $n$
here runs over all indices of the EM basis ($\lambda$, $\mathbf{r}$, $\omega$),
and the $\alpha_n(0)$ correspond to the classical amplitudes obtained when
expressing the laser pulse in the basis defined by these modes. In order to
avoid the explicit propagation of this classical field within a quantum
calculation, the classical and the quantum field can be split in the Hamiltonian
using a time-dependent displacement operator~\cite{Cohen-Tannoudji1987} $T(t) =
e^{\sum_n\alpha_n^*(t) a_n - \alpha_n(t) a_n^\dagger}$, where $\alpha_n(t) =
\alpha_n(0)e^{-i\omega_n t}$. Applying this transformation to the wavefunction,
$|\psi'\rangle = T(t)|\psi\rangle$ adds a (time-dependent) energy shift that
does not affect the dynamics and splits the electric-field term in
\autoref{eq:H_AF_multipolar_dipole} into a classical and a quantum part,
$\hat{\mathbf{E}}(\mathbf{r})' = \hat{\mathbf{E}}(\mathbf{r}) +
\mathbf{E}_{\mathrm{cl}}(\mathbf{r},t)$.

We note that the above properties imply
that within this framework, the action of any incoming laser pulse on the
\emph{full} emitter-cavity system can be described purely by the action of the
medium-supported classical electric field on the emitters, with no additional
explicit driving of any EM modes. This is different to, e.g., input-output
theory, where the EM field is split into modes inside the cavity and free-space
modes outside, and external driving thus affects the cavity modes. Importantly,
$\mathbf{E}_{\mathrm{cl}}(\mathbf{r},t)$ is the field obtained at the position
of the emitter upon propagation of the external laser pulse through the material
structure, i.e., it contains any field enhancement and temporal distortion.
Since the Hamiltonian expressed by the field operators $\ff$ by construction
solves Maxwell's equations in the presence of the material structure,
$\mathbf{E}_{\mathrm{cl}}(\mathbf{r},t)$ can be obtained by simply solving
Maxwell's equations using any classical EM solver without ever expressing the
pulse in the basis of the modes $\ff$. It is important to remember that EM field
observables are also transformed according to
\begin{equation}
\langle\psi|O|\psi\rangle = \langle\psi'|T(t)OT^{\dagger}(t)|\psi'\rangle,
\end{equation}
such that, e.g., $\langle\psi|a_n|\psi\rangle =
\langle\psi'|a_n+\alpha_n(t)|\psi'\rangle$. This takes into account that the
``quantum'' field generated by the laser-emitter interaction interferes with the
classical pulse propagating through the structure, and ensures a correct
description of absorption, coherent scattering, and similar effects.

\subsection{Emitter-centered modes}\label{sec:emitter_centered_modes}
Following~\cite{Buhmann2008Casimir,Hummer2013,Rousseaux2016,Dzsotjan2016,
Castellini2018,Varguet2019Non-hermitian}, we now look for a linear
transformation of the bosonic modes $\ff$ at each frequency in
such a way that in the new basis, only a minimal number of EM modes
couples to the emitters. To this end, we start with the
macroscopic QED Hamiltonian within the multipolar
approach~\autoref{eq:H_MQED_multipolar_full}, with the
emitter-field interaction treated within the long-wavelength
approximation~\autoref{eq:H_AF_multipolar_dipole}. For simplicity
of notation, we assume that the dipole operator of each emitter
only couples to a single field polarization, $\hat{\mathbf{d}}_i =
\hat{\mu}_i \mathbf{n}_i$. Alternatively, the sum over $i$ could
simply be extended to include up to three separate orientations
per emitter. Our goal can then be achieved by defining
\emph{emitter-centered} or \emph{bright} (from the emitter
perspective) EM modes $\hat{B}_i(\omega)$ associated with each
emitter $i$:
\begin{subequations}
\begin{gather}\label{eq:bright_mode}
\hat{B}_{i}(\omega) = \Sum_\lambda \Int\dd^3\mathbf{r}\, \bm{\beta}_{i,\lambda}(\mathbf{r},\omega) \cdot \ff \\
\bm{\beta}_{i,\lambda}(\mathbf{r},\omega) = \frac{\mathbf{n}_i \cdot \mathbf{G}_\lambda(\mathbf{r}_i,\mathbf{r},\omega)}{G_i(\omega)},
\end{gather}
\end{subequations}
where $G_i(\omega)$ is a normalization factor. Using \autoref{eq:f_comm}, the
commutation relations of the operators $\hat{B}_i(\omega)$ reduce to overlap
integrals of their components, $[\hat{B}_i(\omega),\hat{B}_j^\dagger(\omega')] =
S_{ij}(\omega) \delta(\omega-\omega')$, with
\begin{multline}\label{eq:Sij}
    S_{ij}(\omega) = \Sum_\lambda \Int\dd^3\mathbf{r}\, \bm{\beta}_{i,\lambda}^*(\mathbf{r},\omega) \cdot \bm{\beta}_{j,\lambda}(\mathbf{r},\omega) \\
     = \frac{\hbar\omega^2}{\pi\epsilon_0 c^2} \frac{\mathbf{n}_i \cdot \Im\mathbf{G}(\mathbf{r}_i,\mathbf{r}_j,\omega) \cdot \mathbf{n}_j}{G_i(\omega)G_j(\omega)},
\end{multline}
such that the overlap matrix $S_{ij}(\omega)$ at each frequency is real and
symmetric (since $\mathbf{G}(\mathbf{r},\mathbf{r}',\omega) =
\mathbf{G}^T(\mathbf{r}',\mathbf{r},\omega)$). In the above derivation, we have
used the Green's function identity
\begin{equation}\label{eq:GF_integral_identity}
    \sum_\lambda \Int\dd^3\mathbf{s}\, \mathbf{G}_\lambda(\mathbf{r},\mathbf{s},\omega) \cdot \mathbf{G}_\lambda^{*T}(\mathbf{r}',\mathbf{s},\omega) = \frac{\hbar\omega^2}{\pi\epsilon_0 c^2} \Im \mathbf{G}(\mathbf{r},\mathbf{r}',\omega)
\end{equation}
to obtain a compact result~\cite{Scheel2008,Buhmann2012I}. The normalization
factor $G_i(\omega)$ is obtained by requiring that $S_{ii}(\omega) = 1$, so
\begin{equation}\label{eq:bright_mode_normalization}
G_i(\omega) = \sqrt{\frac{\hbar\omega^2}{\pi\epsilon_0 c^2} \mathbf{n}_i \cdot \Im \mathbf{G}(\mathbf{r}_i,\mathbf{r}_i,\omega) \cdot \mathbf{n}_i}.
\end{equation}
We note that the coupling strength $G_i(\omega)$ of the emitter-centered mode
$\hat{B}_i(\omega)$ to emitter $i$ is directly related to the EM spectral
density $J_i(\omega) = [\mu G_i(\omega) / \hbar]^2$ for transition dipole
moment $\mu$~\cite{Novotny2012}. In the regime of weak coupling, i.e., when the
EM environment can be approximated as a Markovian bath, the spontaneous emission
rate at an emitter frequency $\omega_e$ is then given by $2\pi J_i(\omega_e)$.

Since the overlap matrix $S_{ij}(\omega)$ of the modes associated with emitters
$i$ and $j$ is determined by the imaginary part of the Green's function
between the two emitter positions, it follows that the modes
$\hat{B}_i(\omega)$, or equivalently, the coefficient arrays
$\bm{\beta}_i(\mathbf{r},\omega)$, are not orthogonal in general. This can be
resolved by performing an explicit orthogonalization, which is possible as long
as the modes are linearly independent. When this is not the case, linearly
dependent modes can be dropped from the basis until a minimal set is
reached~\cite{Dzsotjan2016}. In the following, we thus assume linear
independence. We can then define new orthonormal modes $\hat{C}_i(\omega)$ as a
linear superposition of the original modes (and vice versa)
\begin{subequations}\label{eq:n_emitter_modes}
\begin{align}
    \hat{C}_i(\omega) &= \sum_{j=1}^N V_{ij}(\omega) \hat{B}_{j}(\omega),\\
    \bm{\chi}_{i,\lambda}(\mathbf{r},\omega) &= \sum_{j=1}^N V_{ij}(\omega) \bm{\beta}_{j,\lambda}(\mathbf{r},\omega),
\end{align}
\end{subequations}
which also implies that $\hat{B}_{i}(\omega) = \sum_{j=1}^N W_{ij}(\omega)
\hat{C}_j(\omega)$, where $\mathbf{W}(\omega) = \mathbf{V}(\omega)^{-1}$. In the
above, the transformation matrix $\mathbf{V}(\omega)$ is chosen such that
$[\hat{C}_i(\omega),\hat{C}_j^\dagger(\omega')] = \delta_{ij}
\delta(\omega-\omega')$, which implies $\mathbf{V}(\omega) \mathbf{S}(\omega)
\mathbf{V}^\dagger(\omega) = \mathbb{1}$. The coefficient matrices
$\mathbf{V}(\omega)$ can be chosen in various ways, corresponding to different
unitary transformations of the same orthonormal basis. We mention two common
approaches here. One consists in performing a Cholesky decomposition of the
overlap matrix, $\mathbf{S}(\omega) = \mathbf{L}(\omega) \mathbf{L}^T(\omega)$,
with $\mathbf{V}(\omega) = \mathbf{L}(\omega)^{-1}$, where $\mathbf{L}(\omega)$
and $\mathbf{L}(\omega)^{-1}$ are lower triangular matrices. This is the result
obtained from Gram-Schmidt orthogonalization, with the advantage that
$\hat{C}_i(\omega)$ only involves $\hat{B}_j(\omega)$ with $j\leq i$ and vice
versa, such that emitter $i$ only couples to the first $i$ photon continua.
Another possibility is given by Löwdin orthogonalization, with
$\mathbf{V}(\omega) = \mathbf{S}(\omega)^{-1/2}$, where the matrix power is
defined as $\mathbf{S}^a = \mathbf{U} \mathbf{\Lambda}^a \mathbf{U}^\dagger$,
with $\mathbf{U}$ ($\mathbf{\Lambda}$) a unitary (diagonal) matrix containing
the eigenvectors (eigenvalues) of $\mathbf{S}$. This approach maximizes the
overlap $[\hat{C}_i(\omega),\hat{B}_i^\dagger(\omega)]$ while ensuring
orthogonality, and can thus be seen as the ``minimal'' correction required to
obtain an orthonormal basis.

Using the orthonormal set of operators $\hat{C}_i(\omega)$, which
are themselves linear superpositions of the operators $\ff$, we
can perform a unitary transformation (separately for each
frequency $\omega$) of the $\ff$ into a basis spanned by the
emitter-centered (or bright) EM modes and an infinite number of
``dark'' modes $\hat{D}_i(\omega)$ that span the orthogonal
subspace and do not couple to the emitters, such that
\begin{equation}
    \label{eq:f_to_C}
    \ff = \sum_{i=1}^N \bm{\chi}_{i,\lambda}^*(\mathbf{r},\omega) \hat{C}_i(\omega)
    + \sum_j \bm{d}_{j,\lambda}^*(\mathbf{r},\omega) \hat{D}_j(\omega) .
\end{equation}
where $\Int\dd^3\mathbf{r}\,\bm{d}_j^*(\mathbf{r},\omega) \cdot
\bm{\chi}_i(\mathbf{r},\omega) = 0$. The Hamiltonian can then be written as
\begin{multline}
    \mathcal{H} = \Int_0^\infty \dd\omega \Bigg[ \sum_{i=1}^N \hbar\omega \hat{C}_i^{\dagger}(\omega) \hat{C}_i(\omega)
    + \sum_{j} \hbar\omega \hat{D}_j^{\dagger}(\omega) \hat{D}_j(\omega) \\
    - \sum_{i,j=1}^N \hat{\mu}_i \left(g_{ij}(\omega) \hat{C}_j(\omega) + \mathrm{H.c.}\right) \Bigg] + \sum_{i=1}^N \hat{H}_i
\end{multline}
where $g_{ij}(\omega) = G_i(\omega) W_{ij}(\omega)$. We note here that
$\mathbf{V}(\omega)$ and thus $\mathbf{W}(\omega)$ and $g_{ij}(\omega)$ can
always be chosen real due to the reality of $S_{ij}(\omega)$, but we here treat
the general case with possibly complex coefficients. Since the dark modes are
decoupled from the rest of the system, they do not affect the dynamics and can
be dropped, giving
\begin{multline}
    \label{eq:emitter_centered_H}
    \mathcal{H} = \sum_{i=1}^N \hat{H}_i
    + \Int_0^\infty \dd\omega \Bigg[ \sum_{i=1}^N \hbar\omega \hat{C}_i^{\dagger}(\omega) \hat{C}_i(\omega) \\
    - \sum_{i,j=1}^N \hat{\mu}_i \left(g_{ij}(\omega) \hat{C}_j(\omega) + \mathrm{H.c.}\right) \Bigg].
\end{multline}
We mention for completeness that if the dark modes are initially excited,
including them might be necessary to fully describe the state of the system. We
have now explicitly constructed a Hamiltonian with only $N$ independent EM modes
$\hat{C}_i(\omega)$ for each frequency $\omega$~\cite{Buhmann2008Casimir,
Hummer2013, Dzsotjan2016, Rousseaux2016, Castellini2018,
Varguet2019Non-hermitian}.

Furthermore, one can obtain an explicit expression
for the electric field operator based on the modes $\hat{C}_i(\omega)$, which to
the best of our knowledge has not been presented previously in the literature.
This is obtained by inserting \autoref{eq:f_to_C} in \autoref{eq:E_MQED}, again
dropping the dark modes $\hat{D}_i(\omega)$, and again using the integral
relation for Green's functions from \autoref{eq:GF_integral_identity}, leading
to
\begin{subequations}\label{eq:e_field_final}
\begin{align}
    \hat{\mathbf{E}}^{(+)}(\mathbf{r}) &= \sum_{i=1}^N \Int_0^\infty \dd\omega\, \mathbf{E}_i(\mathbf{r},\omega) \hat{C}_i(\omega)\\
    \mathbf{E}_i(\mathbf{r},\omega) &= \sum_{j=1}^N V_{ij}^*(\omega) \bm{\mathcal{E}}_j(\mathbf{r},\omega)\\
    \bm{\mathcal{E}}_j(\mathbf{r},\omega) &= \frac{\hbar\omega^2}{\pi\epsilon_0 c^2} \frac{\Im\mathbf{G}(\mathbf{r},\mathbf{r}_j,\omega) \cdot \mathbf{n}_j}{G_j(\omega)}
\end{align}
\end{subequations}
These relations show that we can form explicit photon modes in space at each
frequency by using orthonormal superpositions of the emitter-centered EM modes
$\bm{\mathcal{E}}_j(\mathbf{r},\omega)$. We note that this also provides a
formal construction for the ``emitter-centered'' EM modes, i.e., the modes
created by the operators $\hat{B}_i(\omega)$, with a field profile corresponding
to the imaginary part of the Green's function associated with that emitter.
These modes are well-behaved: they solve the source-free Maxwell equations, are
real everywhere in space, and do not diverge anywhere (here, it should be
remembered that the real part of the Green's function diverges for
$\mathbf{r}=\mathbf{r}'$, while the imaginary part does not). As a simple
example, we can take a single $z$-oriented emitter at the origin in free space.
The procedure used here then gives exactly the $l=1$, $m=0$ spherical Bessel
waves, i.e., the only modes that couple to the emitter when quantizing the field
using spherical coordinates. Finally, we note that it can be verified easily
that inserting \autoref{eq:e_field_final} in \autoref{eq:H_MQED_multipolar_full}
and simplifying leads exactly to \autoref{eq:emitter_centered_H}.

\section{Example}
\begin{figure}[t]
    \includegraphics[width=\linewidth]{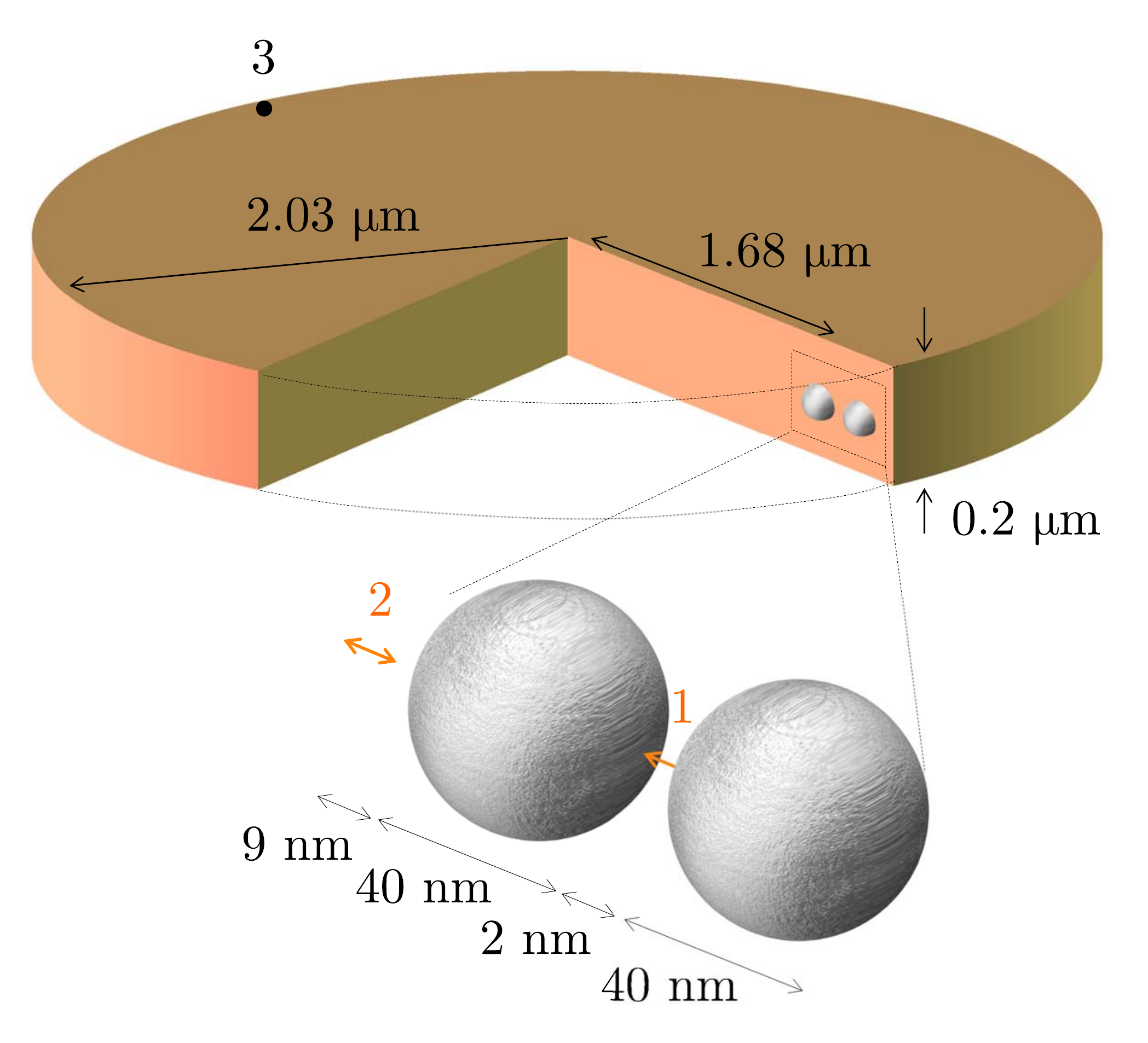}
    \caption{\label{fig:geometry} Sketch of the hybrid metallo-dielectric
    structure we treat: A dielectric microdisk resonator with a metallic dimer
    antenna placed on top. The positions of the two emitters are indicated by
    points $1$ and $2$, while the field evaluation point used later is indicated
    as point $3$.}
\end{figure}

\begin{figure}[t]
    \includegraphics[width=\linewidth]{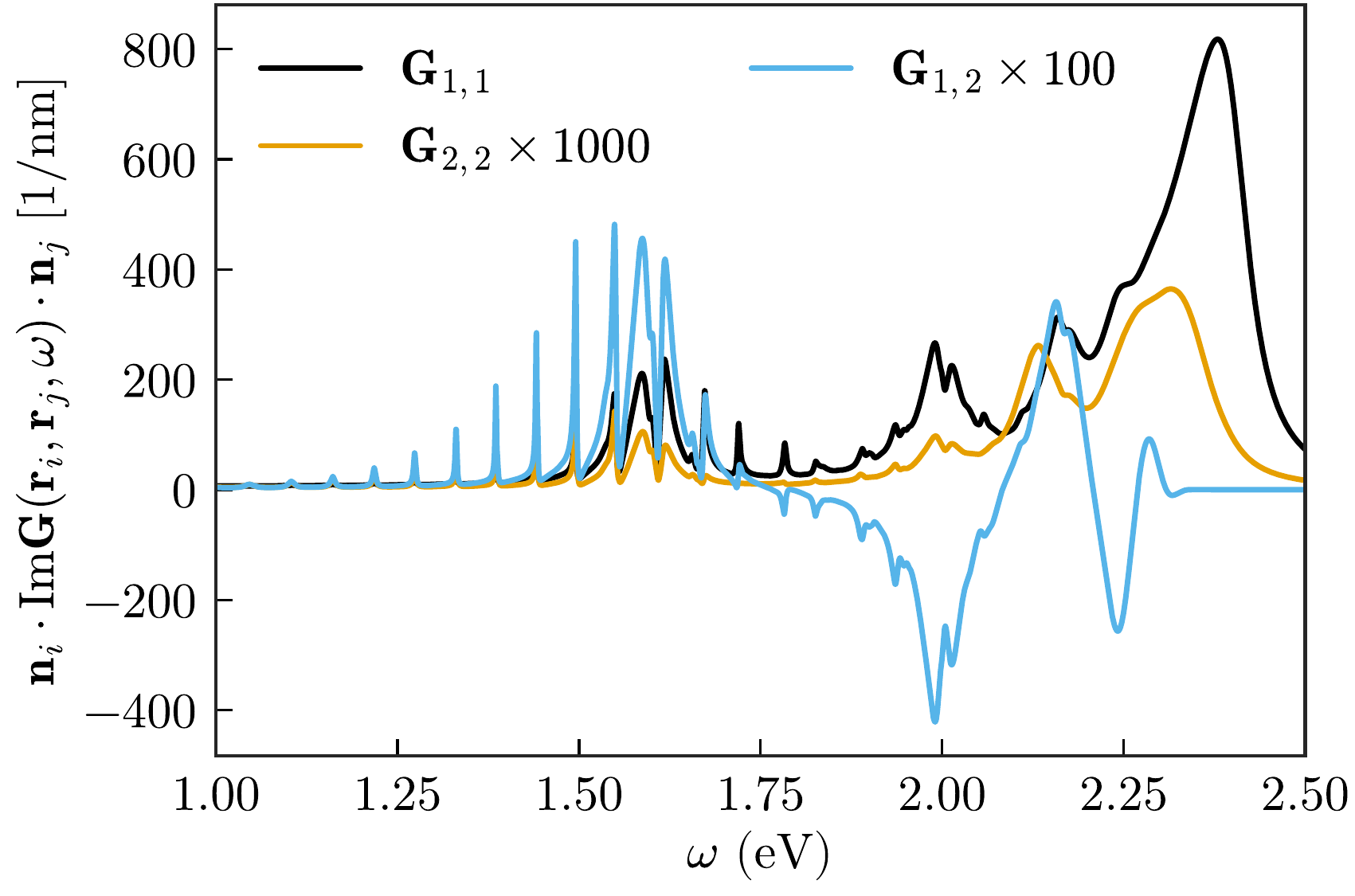}
    \caption{\label{fig:ImG_ij} Green's function factor $\mathbf{n}_i \cdot
    \Im\mathbf{G}(\mathbf{r}_i,\mathbf{r}_j,\omega) \cdot \mathbf{n}_j$
    connecting the two emitters in the geometry of \autoref{fig:geometry}.}
\end{figure}

As an example, here we treat a complex metallodielectric structure, as shown in
\autoref{fig:geometry}. It is composed of a dielectric microdisk resonator
supporting whispering-gallery modes, with a metallic sphere dimer antenna placed
within. The SiN ($\epsilon=4$) disk is similar to that considered in~\cite{Doeleman2016}, with radius 2.03~$\mu$m and height 0.2~$\mu$m.
Two 40 nm diameter Ag (with permittivity taken from~\cite{Rakic1998}) nanospheres
separated by a 2~nm gap are placed 1.68~$\mu$m away from the disk axis.
Two point-dipole emitters modeled
as two-level systems and oriented along the dimer axis are placed in the central
gap of the dimer antenna, and just next to the antenna (labelled as points $1$
and $2$ in \autoref{fig:geometry}). Their transition frequencies are chosen as
$\omega_{e,1} = \omega_{e,2} = 2$~eV, while the dipole transition moments are
$\mu_1 = 0.1$~e\,nm and $\mu_2 = 3$~e\,nm (roughly corresponding to typical
single organic molecules and $J$-aggregates, respectively~\cite{Moll1995}). As
shown in the theory section, the emitter dynamics are then fully determined by
the Green's function between the emitter positions (as also found in
multiple-scattering approaches~\cite{Delga2014}). In \autoref{fig:ImG_ij}, we
show the relevant values $\mathbf{n}_i \cdot
\Im\mathbf{G}(\mathbf{r}_i,\mathbf{r}_j,\omega) \cdot \mathbf{n}_j$, which
clearly reveals the relatively sharp Mie resonances of the dielectric disk,
hybridized with the short-range plasmonic modes of the metallic nanosphere
dimer. In addition, it also shows that the coupling between the two emitters,
i.e., the offdiagonal term with $i=1$, $j=2$, has significant structure and
changes sign several times within the frequency interval.

\begin{figure}[t]
    \includegraphics[width=\linewidth]{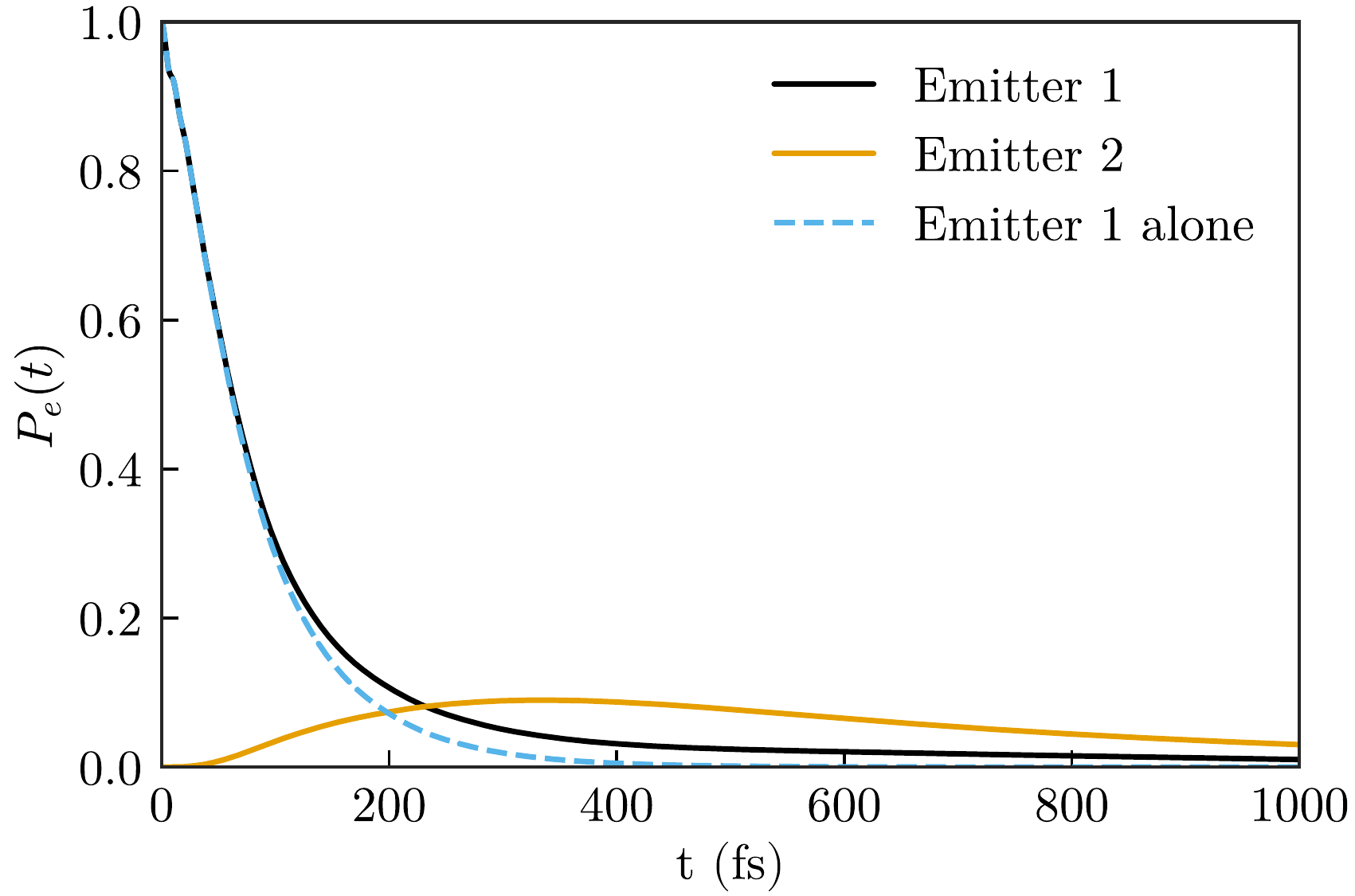}
    \caption{\label{fig:P_emitter} Population of emitters $1$ (black line) and
    $2$ (orange line) for the Wigner-Weisskopf problem with emitter 1 initially
    excited. The blue dashed line shows the dynamics of emitter $1$ if emitter 2
    is not present.}
\end{figure}

\begin{figure}[t]
    \includegraphics[width=\linewidth]{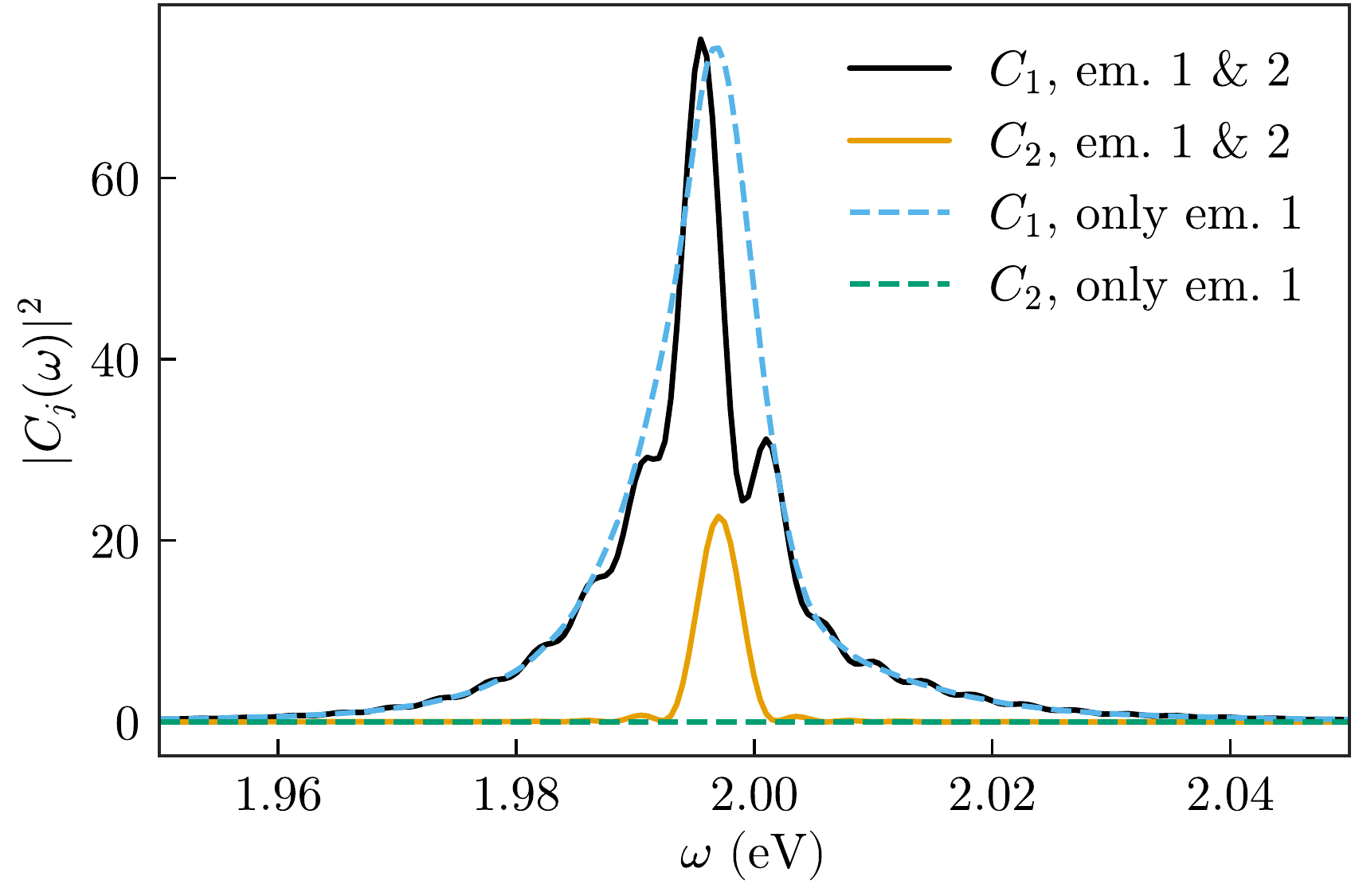}
    \caption{\label{fig:Cj_populations} Population $\langle
    C_{j}^\dagger(\omega) C_{j}(\omega) \rangle$ of EM modes at time $t=1000$~fs
    in the presence of both emitters (solid black and orange lines) and in the
    presence of only emitter $1$ (dashed blue and green lines). Gram-Schmidt
    orthogonalization has been used here, so that emitter $1$ only couples to
    continuum $1$ (i.e., populations for $j=2$ are identically zero when emitter
    $2$ is not present), while emitter $2$ couples to both continua.}
\end{figure}

We now study the dynamics for the Wigner-Weisskopf problem of spontaneous
emission of emitter $1$, i.e., for the case where emitter $1$ is initially in
the excited state, while emitter $2$ is in the ground state and the EM field is
in its vacuum state, i.e., $|\psi(t\!=\!0)\rangle = \sigma_1^{+} |0\rangle$,
where $|0\rangle$ is the global vacuum without any excitations and $\sigma_1^+$
is a Pauli matrix acting on emitter $1$. We also treat the light-matter coupling
within the rotating wave approximation, i.e., we use $\sum_{i,j=1}^N \mu_i
\left(g_{ij}(\omega) \sigma^+ \hat{C}_j(\omega) + \mathrm{H.c.}\right)$ as the
light-matter interaction term. The number of excitations is then conserved, and
the system can be solved easily within the single-excitation subspace by simply
discretizing the photon continua in frequency~\cite{Grynberg2010}. The resulting
emitter dynamics, i.e., the population of the excited states of the emitters,
are shown in \autoref{fig:P_emitter}. This reveals that the EM field emitted by
emitter $1$ due to spontaneous emission is then partially reabsorbed by emitter
$2$. Comparison with the dynamics of emitter $1$ when emitter $2$ is not present
(shown as a dashed blue line in \autoref{fig:P_emitter}) furthermore reveals
that there is also significant transfer of population back from emitter $2$ to
emitter $1$.

\begin{figure}[t]
    \includegraphics[width=\linewidth]{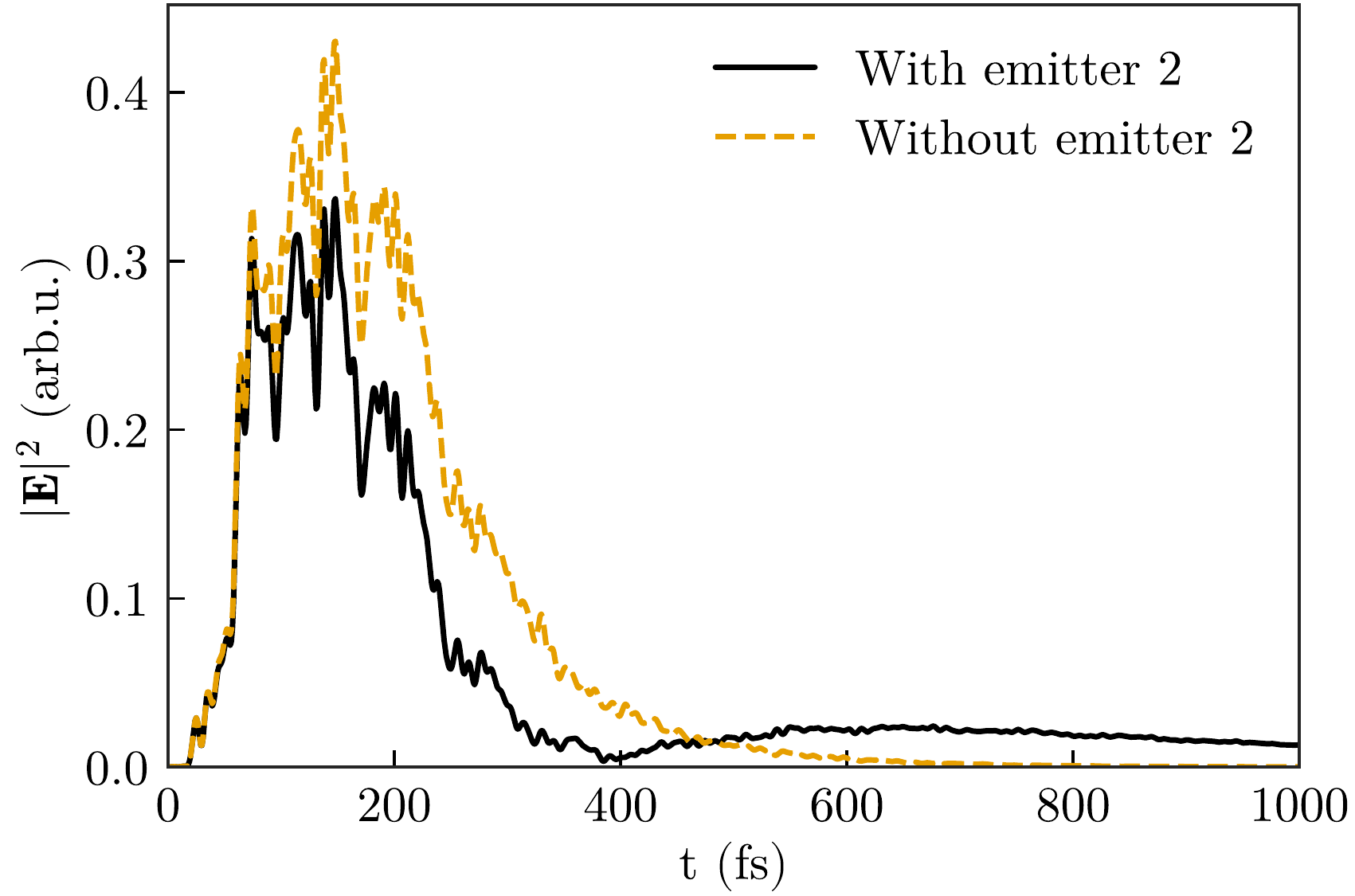}
    \caption{\label{fig:E_field} Time-dependent electric field intensity
    $|\mathbf{E}|^2$ at position $3$ in \autoref{fig:geometry}, for the case
    when both emitters are present (black solid line) and when only emitter $1$
    is present (orange dashed line).}
\end{figure}

The direct access to the photonic modes in this approach provides
interesting insight into, e.g., the photonic mode populations, which are shown
in \autoref{fig:Cj_populations} at the final time considered here, $t=1000$~fs.
As mentioned above, there is some freedom in choosing the orthonormalized
continuum modes $C_j(\omega)$, as any linear superposition of modes at the same
frequency is also an eigenmode of the EM field. We have here chosen the modes
obtained through Gram-Schmidt orthogonalization, which, as discussed above, have
the advantage that emitter $i$ only couples to the first $i$ photon continua. In
particular, emitter $1$ only couples to a single continuum, $\hat{C}_1(\omega)$,
while emitter $2$ couples to the same continuum, and additionally to its ``own''
continuum $\hat{C}_2(\omega)$. This makes the comparison between the case of
having both emitters present or only including emitter $1$ quite direct, as can
be observed in \autoref{fig:Cj_populations}. In particular, any population in
the modes $\hat{C}_2(\omega)$ must come from emitter $2$, after it has in turn
been excited by the photons emitted by emitter $1$ into continuum $1$.

Finally, we also evaluate the electric field in time at a third position
(indicated as point $3$ in \autoref{fig:geometry}), as determined by
\autoref{eq:e_field_final}. This is displayed \autoref{fig:E_field} and shows a
broad initial peak due to the fast initial decay of emitter $1$ (filtered by
propagation through the EM structure, with clear interference effects visible),
and then a longer tail due to the longer-lived emission from both emitters,
which is mostly due to emitter $2$ (which is less strongly coupled to the EM
field) and its back-feeding of population to emitter $1$.

\section{Conclusions}
In this article we have presented a general overview of the application of the
formalism of macroscopic quantum electrodynamics in the context of quantum
nanophotonics. Within this research field, it is often mandatory to describe
from an ab-initio perspective how a collection of quantum emitters interacts
with a nanophotonic structure, which is usually accounted for by utilizing
macroscopic Maxwell equations. Macroscopic QED then needs to combine tools taken
from both quantum optics and classical electromagnetism. After the presentation
of the general formalism and its approximations, we have reviewed in detail the
steps to construct a minimal but complete basis set to analyze the interaction
between an arbitrary dielectric structure and multiple quantum emitters. This
minimal basis set is formed by the so-called emitter-centered modes, such that
all the information regarding the EM environment is encoded into the EM dyadic
Green's function, which can be calculated using standard numerical tools capable
of solving macroscopic Maxwell equations in complex nanophotonic structures. As
a way of example and to show its full potential, in the final part of this
article, we have applied this formalism to solve both the population dynamics
and EM field generation associated to the coupling of two quantum emitters with
a hybrid plasmo-dielectric structure composed of a dielectric microdisk within
which a metallic nanosphere dimer is immersed. We emphasize that this formalism
can be used not only to provide exact solutions to problems in quantum
nanophotonics, but could also serve as a starting point for deriving simpler
models and/or approximated numerical treatments.

% \begin{acknowledgement}
%   Please insert acknowledgments of the assistance of colleagues or similar notes of appreciation here.
% \end{acknowledgement}

\begin{funding}
This work has been funded by the European Research Council
(doi:\doi{10.13039/501100000781}) through grant ERC-2016-StG-714870 and by the
Spanish Ministry for Science, Innovation, and Universities -- Agencia Estatal de
Investigación (doi:\doi{10.13039/501100011033}) through grants
RTI2018-099737-B-I00, PCI2018-093145 (through the QuantERA program of the
European Commission), and MDM-2014-0377 (through the María de Maeztu program for
Units of Excellence in R\&D).
\end{funding}

\bibliographystyle{apsrev4-2}
\nocite{apsrev42Control}
\bibliography{apsrevbibopts,references}

\end{document}